\documentclass[conference,10pt]{IEEEtran}

\usepackage{cite}
\usepackage{graphicx}  
\usepackage{caption}
\usepackage{amsfonts} 
\usepackage{amsmath}
\usepackage{amsthm}
\usepackage{amssymb}

\usepackage{float}
\usepackage{algorithm}  
\usepackage{algpseudocode}  
\usepackage{amsmath}  

\usepackage{booktabs}
\usepackage{array, caption, threeparttable}
\usepackage[font=small]{caption}

\usepackage{graphicx}
\usepackage{subfigure}

\begin{document}
	
\title{Reconfigurable Intelligent Surface Aided Constant-Envelope Wireless Power Transfer}


\author{
	\IEEEauthorblockN{Huiyuan Yang, Xiaojun Yuan,~\IEEEoverridecommandlockouts\IEEEmembership{Senior Member\textrm{,}~IEEE}, Jun Fang\textrm{,}~\IEEEoverridecommandlockouts\IEEEmembership{Senior Member,~IEEE}, and} 
	\IEEEauthorblockN{Ying-Chang Liang\textrm{,}~\IEEEoverridecommandlockouts\IEEEmembership{Fellow,~IEEE}}
	\IEEEauthorblockA{
		\text{National Key Laboratory of Sci. and Tech., University of Electronic Sci. and Tech. of China, Chengdu, China} \\
		Emails: hyyang@std.uestc.edu.cn, \{xjyuan, JunFang\}@uestc.edu.cn, liangyc@ieee.org}
}

\maketitle

\begin{abstract}

By reconfiguring the propagation environment of electromagnetic waves artificially, reconfigurable intelligent surfaces (RISs) have been regarded as a promising and revolutionary hardware technology to improve the energy and spectrum efficiency of wireless networks. In this paper, we study a RIS aided multiuser multiple-input single-output (MISO) wireless power transfer (WPT) system, where the transmitter is equipped with a constant-envelope analog beamformer. We formulate a novel problem to maximize the total received power of all the users by jointly optimizing the beamformer at transmitter and the phase shifts at the RISs, subject to the individual minimum received power constraints of users. We further solve the problem iteratively with a closed-form expression for each step. Numerical results show the performance gain of deploying RIS and the effectiveness of the proposed algorithm.

\end{abstract}


%
\IEEEpeerreviewmaketitle

\section{Introduction}

The proliferation of wireless devices has brought a lot of convenience to our lives, but their limited battery life requires frequent battery replacement/recharging, which is usually costly, sometimes even is infeasible in many critical applications. By powering wireless devices with radio frequency (RF) energy over the air using a dedicated power transmitter, wireless power transfer (WPT) technology provides an attractive solution \cite{bi2015wireless,zeng2017communications}.

The low efficiency of wireless power transfer for users over long distances has been regarded as the performance bottleneck in practical systems \cite{zeng2017communications}. To alleviate this problem, array-based energy beamforming techniques are widely used to obtain a beamforming gain by concentrating the energy of the emitted electromagnetic waves in a narrow spatial angle \cite{liu2018transmit,yang2015throughput,wang2017beamforming,xu2014multiuser}. However, conventional digital beamforming requires that each antenna has its own radio frequency chain, which is costly especially when the antenna number becomes very large \cite{zhang2014massive}. One way to reduce this cost is to apply analog beamforming, in which multiple transmit antennas share a common radio frequency chain with an radio frequency frontend to control the signal amplitude and phase at each antenna \cite{hur2013millimeter,zhang2017multi}. In this paper, we consider constant-envelope analog beamforming, which reduces the frontend to a set of phase shifters (one for each transmit antenna), thereby further reducing the hardware cost \cite{zhang2017multi,zhang2014massive,mohammed2013per}.

Another promising way to increase the efficiency of wireless power transfer is to place reconfigurable intelligent surfaces (RISs) in the wireless propagation environment so as to recofigure the propagation environment of electromagnetic waves artificially. RISs is a kind of programmable and reconfigurable passive meta-surfaces which are able to change the signal transmission direction \cite{yu2019miso}. Typically, a RIS is a planar array composed of low-cost passive elements, e.g., printed dipoles, where each of the elements can independently reflect impinging electromagnetic waves with a controllable phase shift, thus collaboratively changing the reflcted signal propagation \cite{wu2019intelligent,yuan2020reconfigurable}. In this way, IRSs are able to create favorable wireless propagation environments through intelligent placement and passive beamforming \cite{yuan2020reconfigurable}, so as to increase the efficiency of wireless power transfer.

 \begin{figure}[h]
	 \centerline{\includegraphics[width=0.73\columnwidth]{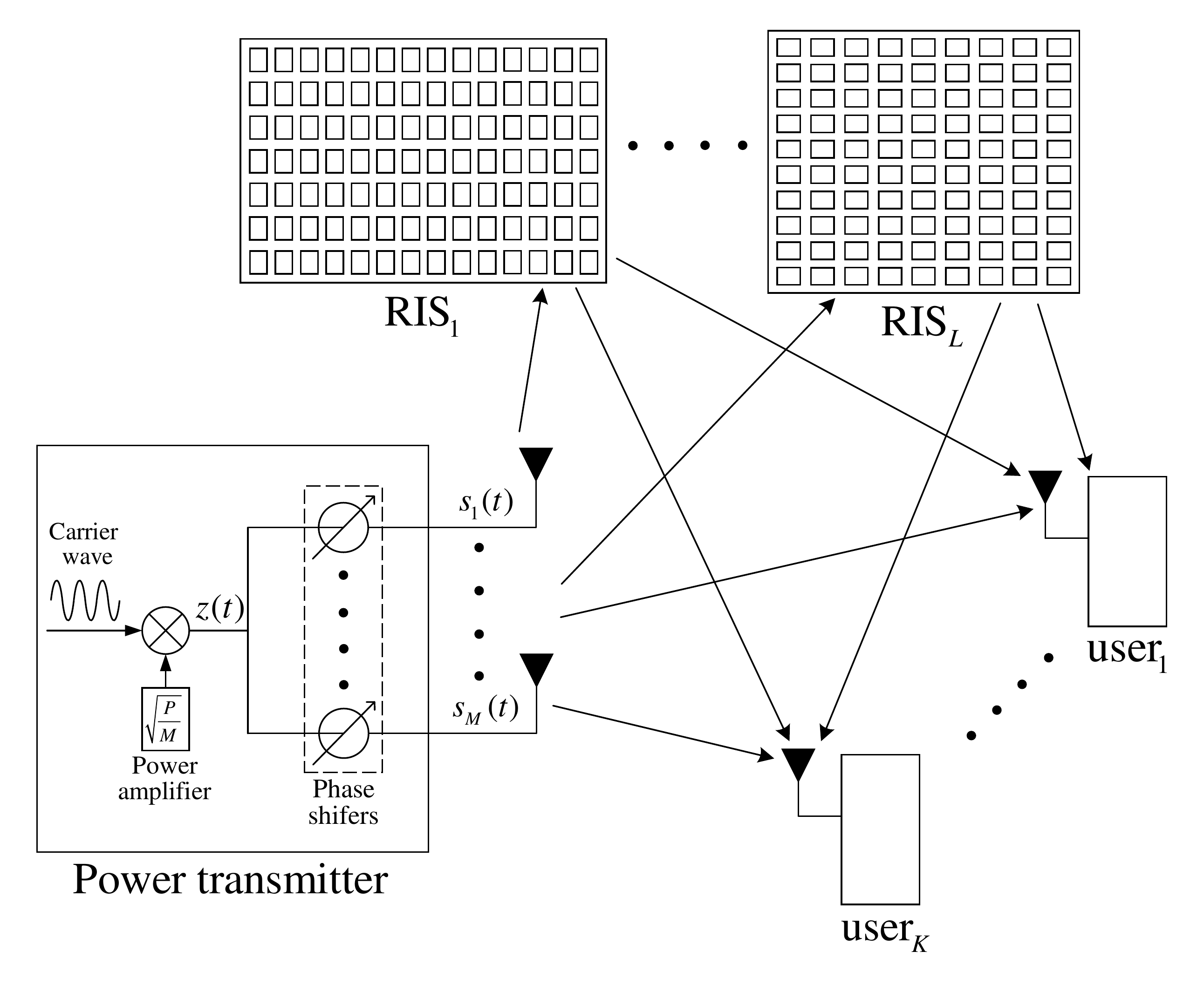}}
	\caption{ RIS aided wireless power transfer in a multiuser MISO system with a constant-envelope analog beamformer at the multi-antenna transmitter.
	}
    \vspace{-0.1cm}
\end{figure}

In this paper, we study a multiuser multiple-input single-output (MISO) wireless power transfer system which is aided by several RISs. As shown in Fig. 1, the system consists of $L$ RISs, $K$ single-antenna receivers and a transmitter equipped with $M$ antennas. Following the idea of analog beamforming, all the antennas at the transmitter share a common RF chain.

A typical application of the considered system is to charge the mobile devices in the office area. In this scenario, the mobile devices cannot stay in the charging area for long periods of time, so the charging service of each device must be completed in a relatively short time. Further, different devices may have different requirements on charging speed due to the difference of remaining power or other reasons. This inspires us to propose an optimization problem that takes into account both energy efficiency and user fairness, i.e., to maximize the total received power of all the users subject to the individual minimum received power constraints of users and the constant-envelope constraints, where the constant-envelope constraints include two parts: 1) the signals radiated by all the transmitting antennas are required to have the same constant amplitude; 2) each element of the RIS generates a fixed energy gain on the signal it reflects. To solve this problem, we propose a low-complexity iterative algorithm with a closed-form expression for each step, based on the successive convex approximation (SCA) method and the alternating direction method of multipliers (ADMM) method. Numerical results show the performance gain of deploying RIS and the effectiveness of the proposed algorithm.

\section{System Model And Problem Formulation}

\subsection{System Model}

Consider a multiuser MISO wireless power transfer system assisted by a number of reconfigurable intelligent surfaces, where a base station (BS) with $M$ antennas transfers power to $K$ single-antenna users simultaneously. The system consists of $L$ RISs, where the $l$-th RIS, denoted by $\textrm{RIS}_l$, has $N_l$ elements, $\forall l$. Assume that the power of the signals reflected by the RIS two or more times is negligible and thus ignored due to the significant path loss. In addition, all the channels in this paper are assumed to be quasi-static flat-fading channels.

The baseband equivalent channels of the BS-user link, the $\textrm{RIS}_l$-user link and the BS-$\textrm{RIS}_l$ link are denoted by $\boldsymbol{H}_{d} \in \mathbb{C}^{K \times M}$, $\boldsymbol{H}_{r,l} \in \mathbb{C}^{K \times N_l}$ and $\boldsymbol{S}_l \in \mathbb{C}^{N_l \times M}$, respectively, where $\mathbb{C}^{a \times b}$ denotes the space of $a \times b$ complex-valued matrices. Let $N=\sum_{l=1}^{L}N_l$ be the total number of reflecting elements. The baseband equivalent channels of the RIS-user link and the BS-RIS link are denoted by $\boldsymbol{H}_{r} \in \mathbb{C}^{K \times N}$ and $\boldsymbol{S} \in \mathbb{C}^{N \times M}$, respectively, where $\boldsymbol{H}_r=[\boldsymbol{H}_{r,1},\dots,\boldsymbol{H}_{r,L}]$, $\boldsymbol{S}=[\boldsymbol{S}_1^T,\dots,\boldsymbol{S}_L^T]^T$ and $\boldsymbol{D}^T$ denotes the transpose of matrix $\boldsymbol{D}$. We further assume that the channel matrices $\boldsymbol{H}_r$, $\boldsymbol{H}_d$ and $\boldsymbol{S}$ are perfectly known. 

The gain vector of the $l$-th RIS is denoted by $\boldsymbol{\phi}_l=[\beta_{l,1} e^{j\theta_{l,1}},\ldots,\beta_{l,N_l} e^{j\theta_{l,N_l}}]^T$, where $j$ denotes the imaginary unit, $\beta_{l,n} \in [0,1]$ and $\theta_{l,n} \in [-\pi,\pi)$ are the amplitude and phase of the reflection coefficient of the $n$-th element on the RIS, respectively. We assume that the amplitudes of the reflection coefficients are known and unchangeable. Let $\boldsymbol{\phi}=[\boldsymbol{\phi}_1^T,\dots,\boldsymbol{\phi}_L^T]^T=[\beta_1 e^{j\theta_{1}},\ldots,\beta_n e^{j\theta_{n}},\ldots,\beta_N e^{j\theta_{N}}]^T$ be the gain vector of all the RISs. Then, the multiuser MISO channel matrix $\boldsymbol{H} \in \mathbb{C}^{K \times M}$ can be represented as
 \begin{equation} 
\boldsymbol{H}=\boldsymbol{H}_{r} \boldsymbol{\Phi} \boldsymbol{S} + \boldsymbol{H}_{d},
\end{equation}
where $\boldsymbol{\Phi} = \textrm{diag}(\boldsymbol{\phi})$ and diag$(\boldsymbol{d})$ denotes a diagonal matrix with each diagonal element being the corresponding element in $\boldsymbol{d}$. For the beauty of the subsequent formulas in the paper, let phase-shift vector $\boldsymbol{v}=[e^{j\theta_{1}},\ldots, e^{j\theta_{N}}]^H$, where $\boldsymbol{d}^H$ denote the conjugate transpose of vector $\boldsymbol{d}$. Note that $\boldsymbol{\Phi}=\boldsymbol{\Psi}\boldsymbol{\Lambda}$, where $\boldsymbol{\Psi}=\textrm{diag}\left(\boldsymbol{v}^{\star}\right)$, $\boldsymbol{\Lambda}=\textrm{diag}\left([\beta_1,\ldots,\beta_N]\right)$ and $\boldsymbol{d}^{\star}$ is the conjugate of $\boldsymbol{d}$. Then, (1) can be rewritten as
 \begin{equation} 
\boldsymbol{H}=\boldsymbol{H}_{r} \boldsymbol{\Psi} \boldsymbol{G} + \boldsymbol{H}_{d},
\end{equation}
where $\boldsymbol{G}=\boldsymbol{\Lambda}\boldsymbol{S}$ is the modified baseband equivalent channel matrix of the BS-RIS link.

We assume that all the $M$ antennas of the BS share a common RF chain to reduce the implementation cost. The RF signal is generated as 
 \begin{equation} 
z(t)=\sqrt{\frac{P}{M}} e^{j 2\pi f_c t},
\end{equation}
where $f_c$ is the carrier frequency and $P$ is the total transmission power at the BS. Then, $z(t)$ goes through the phase shifter at antenna $m$, yielding 
 \begin{equation} 
s_m (t)=z(t)e^{j\alpha_m}=x_m e^{j 2\pi f_c t}, m=1,\dots,M,
\end{equation}
where $x_m=\sqrt{P/M} e^{j \alpha_m}$ and $\alpha_m \in [-\pi,\pi)$ is the phase-shift at the $m$-th antenna.  At time $t$, the received signal vector of all the $K$ users is given by
 \begin{equation} 
\boldsymbol{y}(t)=\boldsymbol{H}\boldsymbol{s}(t),
\end{equation}
where $\boldsymbol{s}(t) = [s_1(t),s_2(t),\dots,s_M(t)]^T \in \mathbb{C}^{M \times 1}$ with the element $s_m(t)$ given by (4) and $\boldsymbol{y}(t) = [y_1(t),y_2(t),\dots,y_K(t)]^T \in \mathbb{C}^{K \times 1} $ with $y_k(t)$ being the received signal of user $k$. In (5), we assume that the additive noise at the receiver is sufficiently weak compared with the received signal and thus can be ignored for energy harvesting \cite{zhang2017multi}.

Each user aims to harvest energy from the radio frequency signal $y_k(t)$ it received. According to the law of energy conservation, the power of the received radio frequency band signal is proportional to the received baseband power \cite{zhang2017multi}. Thus the received power of user $k$ is given by
 \begin{equation} 
Q_k=\frac{\eta}{T} \int_{0}^{T} \left|y_k(t)\right|^2 dt =\eta\left|\boldsymbol{h}_k^H\boldsymbol{x}\right|^2,
\end{equation}
where $\eta \in \left[0,1\right]$ is the energy conversion efficiency at the user receivers, $T=1/f_c$ is the period of the carrier signal, $\boldsymbol{h}_{k}^H$ is the $k$-th row of the channel matrix $\boldsymbol{H}$, and $\boldsymbol{x}=[x_1,\dots,x_M]^T$. In practice, $\eta$ is determined by both the transmit signal and the receiver structure. For simplicity, we assume $\eta=1$ in the sequel.

\subsection{Problem Formulation}
We consider the optimization problem of maximizing the total received power subject to the minimum received power constraints, formulated as

\begin{alignat}{2}
\textrm{(P1):} \quad \max_{\boldsymbol{\Psi},\boldsymbol{x}}& \
\sum_{k=1}^{K} \left(Q_k= \left|\boldsymbol{h}_{r,k}^{H} \boldsymbol{\Psi}\boldsymbol{G}\boldsymbol{x}+\boldsymbol{h}_{d,k}^{H}\boldsymbol{x} \right|^2\right) & \\
\mbox{s.t.}\quad
&[\boldsymbol{\Psi}]_{n,n}=e^{j\theta_{n}}, \theta_n \in [-\pi,\pi), n=1,\dots,N,\\
&[\boldsymbol{x}]_m=\sqrt{\frac{P}{M}}e^{j\alpha_m}, \alpha_m \in [-\pi,\pi), m=1,\dots,M,\\
&\left|\boldsymbol{h}_{r,k}^{H} \boldsymbol{\Psi}\boldsymbol{G}\boldsymbol{x}+\boldsymbol{h}_{d,k}^{H}\boldsymbol{x} \right|^2 \geq p_k,k=1,\dots,K,
\end{alignat}
where $[\boldsymbol{d}]_n$ denotes the $n$-th element in vector $\boldsymbol{d}$, $[\boldsymbol{D}]_{n,n}$ denotes the element of matrix $\boldsymbol{D}$ in row $n$, column $n$, $p_k$ refers to the minimum received power allowed for user $k$, $\boldsymbol{h}_{r,k}^{H}$ and $\boldsymbol{h}_{d,k}^{H}$ represent the $k$-th rows of matrix $\boldsymbol{H}_r$ and $\boldsymbol{H_d}$, respectively.

Due to the constant-envelope constraints and the non-concave objective funtion with respect to $\boldsymbol{\Psi}$ and $\boldsymbol{x}$, (P1) is non-convex and thus is in general difficult to solve exactly. however, if we fix $\boldsymbol{\Psi}$ (or $\boldsymbol{x}$), (P1) is a non-convex quadratically constrained quadratic program (QCQP) with respect to $\boldsymbol{x}$ (or $\boldsymbol{\Psi}$). This observation inspires us to leverage alternating optimization technique, which is an iterative procedure for maximizing the objective function by alternating maximizations over the individual subsets of the variables.

\section{Sum Power Maximization with Minimum Received Power Constraints}

In this section, we propose a suboptimal solution based on alternating optimization to solve (P1), namely, SPMC-SCA-ADMM algorithm, by leveraging the SCA and ADMM techniques.

\subsection{Optimization of $\boldsymbol{x}$ for a fixed $\boldsymbol{\Psi}$}

With $\boldsymbol{\Psi}$ fixed, (P1) can be written as follows:
\begin{alignat}{2}
\textrm{(P2):} \quad \max_{\boldsymbol{x}} \quad & 
\|\boldsymbol{H}\boldsymbol{x}\|^2 & \\
\mbox{s.t.}\quad
&\left|\boldsymbol{h}_k^H\boldsymbol{x}\right|^2 \geq p_k,k=1,\dots,K,\\
&\left|[\boldsymbol{x}]_m\right|=\sqrt{\frac{P}{M}}, m=1,\dots,M,
\end{alignat}
where $\|\boldsymbol{d}\|$ represents the Frobenius norm of vector $\boldsymbol{d}$.

To tackle this problem, we first use the SCA framework to transform the problem form, and then use the ADMM algorithm to solve it. To apply the SCA method, we need to find a suitable lower bound of $\|\boldsymbol{H}\boldsymbol{x}\|^2$. To do so, we expand $\|\boldsymbol{H}\boldsymbol{x}\|^2$ at a feasible point $\hat{\boldsymbol{x}}$ to obtain a linear lower bound as follow
\begin{equation}
\|\boldsymbol{H}\boldsymbol{x}\|^2 \geq 2 \text{Re}\left\{\hat{\boldsymbol{x}}^H\boldsymbol{H}^H\boldsymbol{H}\boldsymbol{x}\right\}-\hat{\boldsymbol{x}}^H\boldsymbol{H}^H\boldsymbol{H}\hat{\boldsymbol{x}},
\end{equation}
where the equality holds at point $\boldsymbol{x}=\hat{\boldsymbol{x}}$. Next, we use the ADMM algorithm to maximize this lower bound under the constraints of (12) and (13). The corresponding optimization problem is as follows
\begin{alignat}{2}
\textrm{(P3):} \quad \max_{\boldsymbol{x}} \quad & \text{Re}\left\{\hat{\boldsymbol{x}}^H\boldsymbol{H}^H\boldsymbol{H}\boldsymbol{x}\right\}& \\
\mbox{s.t.}\quad
&\textrm{(12), (13)}.
\end{alignat}
The above problem can be written in the following form
\begin{alignat}{2}
\textrm{(P4):} \quad \min_{\boldsymbol{x},\left\{\boldsymbol{e}_k \right\}_{k=1}^{K}} &
\text{Re}\left\{-\hat{\boldsymbol{x}}^H\boldsymbol{H}^H\boldsymbol{H}\boldsymbol{x}\right\} \\
\mbox{s.t.}\quad
&\left|\boldsymbol{h}_k^H \boldsymbol{e}_k\right|^2 \geq p_k, k=1,\dots,K,\\
&\left|\left[\boldsymbol{x}\right]_m\right|=\sqrt{\frac{P}{M}}, m=1,\dots,M,\\
&\boldsymbol{e}_k=\boldsymbol{x}, k=1,\dots,K.
\end{alignat}
Define the feasible region of constraint (18) as $\mathcal{G}$, whose indicator function is given by 
\begin{equation}
\mathbb{I}_{\mathcal{G}}\left(\{\boldsymbol{e}_k\}_{k=1}^{K}\right)=\left\{\begin{array}{ll}
0, & \text { if } \{\boldsymbol{e}_k\}_{k=1}^{K} \in \mathcal{G}, \\
+\infty, & \text { otherwise }.
\end{array}\right.
\end{equation}
Similarly, define the feasible region of constraint (19) as $\mathcal{H}$, and its indicator function as
\begin{equation}
\mathbb{I}_{\mathcal{H}}(\boldsymbol{x})=\left\{\begin{array}{ll}
0, & \text { if } \boldsymbol{x} \in \mathcal{H}, \\
+\infty, & \text { otherwise }.
\end{array}\right.
\end{equation}
Then, we obtain the equivalent ADMM form of (P4) as
\begin{alignat}{2}
\textrm{(P5):}  \min_{\boldsymbol{x},\left\{\boldsymbol{e}_k \right\}_{k=1}^{K}}  
&\text{Re}\left\{-\hat{\boldsymbol{x}}^H\boldsymbol{H}^H\boldsymbol{H}\boldsymbol{x}\right\} + \mathbb{I}_{\mathcal{G}}\left(\{\boldsymbol{e}_k\}_{k=1}^{K}\right) + \mathbb{I}_{\mathcal{H}}(\boldsymbol{x})\\
\mbox{s.t.}\quad 
&\textrm{(20)}.
\end{alignat}
The augmented Lagrangian of (P5) can be formulated as
\begin{equation}
\begin{split}
\mathcal{L}_{\rho}\left(\boldsymbol{x},\left\{\boldsymbol{e}_k \right\}_{k=1}^{K}, \left\{\boldsymbol{u}_k \right\}_{k=1}^{K}\right)
=\text{Re}\left\{-\hat{\boldsymbol{x}}^H\boldsymbol{H}^H\boldsymbol{H}\boldsymbol{x}\right\} +\\ \mathbb{I}_{\mathcal{G}}\left(\{\boldsymbol{e}_k\}_{k=1}^{K}\right) + \mathbb{I}_{\mathcal{H}}(\boldsymbol{x}) 
+ \rho \sum_{k=1}^K \left\|\boldsymbol{e}_k-\boldsymbol{x}+\boldsymbol{u}_k\right\|^2,
\end{split}
\end{equation}
where $\rho \textgreater 0$ is the penalty parameter, $\{\boldsymbol{u}_k\}_{k=1}^K$ are the scaled dual variables. Applying the ADMM method, we update the global variable $\boldsymbol{x}$, the local variables $\left\{\boldsymbol{e}_k\right\}_{k=1}^{K}$ and the scaled dual variables $\left\{\boldsymbol{u}_k \right\}_{k=1}^{K}$ alternatively.

In the $i$-th iteration, given $\boldsymbol{x}^{(i)},\{\boldsymbol{e}_k^{(i)}\}_{k=1}^{K}$ and $\{\boldsymbol{u}_k^{(i)}\}_{k=1}^{K}$, we update each of the above variables as follows.

\paragraph{Update $\boldsymbol{x}$}
The subproblem for updating the global variables $\boldsymbol{x}$ is expressed as
\begin{equation}
\begin{split}
\boldsymbol{x}^{(i+1)}=&\mathop{\arg\min}_{\boldsymbol{x}} \ \mathcal{L}_{\rho}\left(\boldsymbol{x},\left\{\boldsymbol{e}_k^{(i)}\right\}_{k=1}^{K},\left\{\boldsymbol{u}_k^{(i)}\right\}_{k=1}^{K}\right),\\
=&\mathop{\arg\min}_{\boldsymbol{x}} \
\mathbb{I}_{\mathcal{H}}(\boldsymbol{x})+ K\rho\boldsymbol{x}^H\boldsymbol{x}-\\
&\text{Re}\left\{\left(\hat{\boldsymbol{x}}^H\boldsymbol{H}^H\boldsymbol{H} +2\rho \sum_{k=1}^K\left(\boldsymbol{u}^{(i)}_k+\boldsymbol{e}^{(i)}_k\right)^H\right)\boldsymbol{x}\right\}.
\end{split}
\end{equation}
Since $\boldsymbol{x}^H\boldsymbol{x}$ is a constant under the constraint (19), it is easy to see that the optimum is given by
\begin{equation}
\begin{split}
\boldsymbol{x}^{(i+1)}
=&\sqrt{\frac{P}{M}} \textrm{exp}\left(j\textrm{arg}\left(\boldsymbol{H}^H\boldsymbol{H}\hat{\boldsymbol{x}}+
2\rho\sum_{k=1}^{K}\left(\boldsymbol{u}^{(i)}_k+\boldsymbol{e}^{(i)}_k\right)\right)\right),
\end{split}
\end{equation}
where $\textrm{arg}(\boldsymbol{d})$ is a vector with each element is the argument of the corresponding element in complex vector $\boldsymbol{d}$ and $\textrm{exp}\left(\boldsymbol{d}\right)=[e^{[\boldsymbol{d}]_1},\dots,e^{[\boldsymbol{d}]_N}]^T$.

\paragraph{Update $\left\{\boldsymbol{e}_k\right\}_{k=1}^{K}$}
The subproblem for updating the local variables $\left\{\boldsymbol{e}_k\right\}_{k=1}^{K}$ is expressed as
\begin{equation}
\begin{split}
\left\{\boldsymbol{e}^{(i+1)}_k\right\}_{k=1}^{K}
=&\mathop{\arg\min}_{\left\{\boldsymbol{e}_k\right\}_{k=1}^{K}} \ \mathcal{L}_{\rho}\left(\boldsymbol{x}^{(i+1)},\left\{\boldsymbol{e}_k\right\}_{k=1}^{K},\left\{\boldsymbol{u}_k^{(i)}\right\}_{k=1}^{K}\right),
\end{split}
\end{equation}
which can be rewritten as
\begin{alignat}{2}
\textrm{(P6):} \quad \min_{\left\{\boldsymbol{e}_k \right\}_{k=1}^{K}}\quad &
\sum_{k=1}^K \left\|\boldsymbol{e}_k-\boldsymbol{x}^{(i+1)}+\boldsymbol{u}^{(i)}_k\right\|^2 \\
\mbox{s.t.}\quad
&\left|\boldsymbol{h}_k^H \boldsymbol{e}_k\right|^2 \geq p_k, k=1,\dots,K.
\end{alignat}
Since in the above problem, the optimizations of the elements in set $\left\{\boldsymbol{e}_k\right\}_{k=1}^{K}$ are decoupled and these optimization problems have the same form, we only need to study one of them as follows
\begin{alignat}{2}
\textrm{(P7):} \quad \min_{\boldsymbol{e}_k}\quad & \left\|\boldsymbol{e}_k-\boldsymbol{x}^{(i+1)}+\boldsymbol{u}^{(i)}_k\right\|^2 \\
\mbox{s.t.}\quad
&\left|\boldsymbol{h}_k^H \boldsymbol{e}_k\right|^2 \geq p_k.
\end{alignat}
For this particular objective function, which is the Euclidean distance from $\boldsymbol{e}_k$ to $\boldsymbol{x}^{(i+1)}-\boldsymbol{u}_k^{(i)}$, the optimum must be on the edge of the feasible region if it is not  $\boldsymbol{x}^{(i+1)}-\boldsymbol{u}_k^{(i)}$. Thus, to tackle (P7), we first check whether $\boldsymbol{x}^{(i+1)}-\boldsymbol{u}_k^{(i)}$ is feasible. If yes, let $\boldsymbol{e}^{(i+1)}_k=\boldsymbol{x}^{(i+1)}-\boldsymbol{u}_k^{(i)}$. Otherwise, solve the following optimization problem instead:
\begin{alignat}{2}
\textrm{(P8):} \quad \min_{\boldsymbol{e}_k}\quad & \left\|\boldsymbol{e}_k-\boldsymbol{x}^{(i+1)}+\boldsymbol{u}^{(i)}_k\right\|^2 \\
\mbox{s.t.}\quad
&\left|\boldsymbol{h}_k^H \boldsymbol{e}_k\right| = \sqrt{p_k}.
\end{alignat}
The constraint (34) can be written as a linear constraint with an unknown phase $\nu$:
\begin{equation}
\boldsymbol{h}_k^H\boldsymbol{e}_k=\sqrt{p_k}e^{j\nu}.
\end{equation}
Suppose we know $\nu$. (P8) becomes a projection onto an affine subspace \cite{huang2016consensus}, whose solution is given by
\begin{equation}
\boldsymbol{e}_k=\boldsymbol{x}^{(i+1)}-\boldsymbol{u}_k^{(i)}+
\frac{\sqrt{p_k}e^{j\nu}-\boldsymbol{h}_k^H \left(\boldsymbol{x}^{(i+1)}-\boldsymbol{u}_k^{(i)}\right)}{\left\|\boldsymbol{h}_k\right\|^2}\boldsymbol{h}_k.
\end{equation}
Plugging this intermediate solution to the objective function of (P8), we see that the minimum is attained if we set $\nu=\textrm{arg}\left(\boldsymbol{h}_k^H \left(\boldsymbol{x}^{(i+1)}-\boldsymbol{u}_k^{(i)}\right)\right)$. Plugging this to (36), we obtain
\begin{equation}
\begin{split}
\boldsymbol{e}^{(i+1)}_k=&\boldsymbol{x}^{(i+1)}-\boldsymbol{u}_k^{(i)}+\\
&\frac{\sqrt{p_k}-\left|\boldsymbol{h}_k^H \left(\boldsymbol{x}^{(i+1)}-\boldsymbol{u}_k^{(i)}\right)\right|}{\left\|\boldsymbol{h}_k\right\|^2  \left|\boldsymbol{h}_k^H \left(\boldsymbol{x}^{(i+1)}-\boldsymbol{u}_k^{(i)}\right)\right|}\boldsymbol{h}_k \boldsymbol{h}_k^H  \left(\boldsymbol{x}^{(i+1)}-\boldsymbol{u}_k^{(i)}\right).
\end{split}
\end{equation}
To summarize, for $k=1,\dots,K$,
\begin{equation}
\begin{cases}
\boldsymbol{e}_k^{(i+1)}=\boldsymbol{x}^{(i+1)}-\boldsymbol{u}_k^{(i)};& \textrm{if} \left|\boldsymbol{h}_k^H\left(\boldsymbol{x}^{(i+1)}-\boldsymbol{u}_k^{(i)}\right)\right|^2 \geq p_k,\\
(37);& \textrm{otherwise}.
\end{cases}
\end{equation}
\paragraph{Update  $\left\{\boldsymbol{u}_k\right\}_{k=1}^{K}$}
According to the ADMM method, the update formulas for scaled dual variables $\left\{\boldsymbol{u}_k\right\}_{k=1}^{K}$ are shown bellow:
\begin{equation}
\boldsymbol{u}_k^{(i+1)}=\boldsymbol{u}_k^{(i)}+\boldsymbol{e}_k^{(i+1)}-\boldsymbol{x}^{(i+1)},\ k=1,\dots,K.
\end{equation}

Iterating using (27), (38) and (39), we finally obtain a suboptimal solution of (P3).

\subsection{Optimization of $\boldsymbol{\Psi}$ for a fixed $\boldsymbol{x}$}

With $\boldsymbol{x}$ fixed, we can rewrite (P1) as
\begin{alignat}{2}
\textrm{(P9):} \quad \max_{\boldsymbol{v}} \quad & 
\sum_{k=1}^{K} \left|\boldsymbol{v}^H\boldsymbol{c}_k+a_k\right|^2& \\
\mbox{s.t.}\quad
&\left|[\boldsymbol{v}]_n\right|=1, n=1,\dots,N,\\
&\left|\boldsymbol{v}^H\boldsymbol{c}_k+a_k\right|^2 \geq p_k,k=1,\dots,K,
\end{alignat}
where $\boldsymbol{c}_k=\textrm{diag}(\boldsymbol{h}_{r,k}^{\star})\boldsymbol{G}\boldsymbol{x}$ and $a_k=\boldsymbol{h}_{d,k}^H\boldsymbol{x}$. By introducing an auxiliary variable $t$, (P9) can be equivalently written as
\begin{alignat}{2}
\textrm{(P10):} \quad \max_{\boldsymbol{b}} \quad & 
\boldsymbol{b}^H\boldsymbol{L}\boldsymbol{b}& \\
\mbox{s.t.}\quad
&\left|\boldsymbol{l}_k^H\boldsymbol{b}\right|^2 \geq p_k,k=1,\dots,K,\\
&\left|\left[\boldsymbol{b}\right]_n\right|=1, n=1,\dots,N+1,
\end{alignat}
where 
\begin{equation} 
\boldsymbol{l}_k=
\begin{bmatrix}
\boldsymbol{c}_k\\
a_k
\end{bmatrix},
\quad
\boldsymbol{L}=\sum_{k=1}^K \boldsymbol{l}_k \boldsymbol{l}_k^H
\ \textrm{and}\
\boldsymbol{b}=
\begin{bmatrix}
t\boldsymbol{v}\\
t
\end{bmatrix}.
\end{equation}

Again, we expand $\boldsymbol{b}^H\boldsymbol{L}\boldsymbol{b}$ at feasible point $\hat{\boldsymbol{b}}$ to obtain a linear lower bound of it as follow
\begin{equation}
\boldsymbol{b}^H\boldsymbol{L}\boldsymbol{b} \geq 2 \text{Re}\left\{\hat{\boldsymbol{b}}^H\boldsymbol{L}\boldsymbol{b}\right\}-\hat{\boldsymbol{b}}^H\boldsymbol{L}\hat{\boldsymbol{b}},
\end{equation}
where the equality holds at point $\boldsymbol{b}=\hat{\boldsymbol{b}}$. Next, we use the ADMM algorithm to maximize this lower bound under the constraints of (44) and (45). The corresponding optimization problem is as follows
\begin{alignat}{2}
\textrm{(P11):} \quad \max_{\boldsymbol{b}} \quad & 
\text{Re}\left\{\hat{\boldsymbol{b}}^H\boldsymbol{L}\boldsymbol{b}\right\}& \\
\mbox{s.t.}\quad
&\textrm{(44), (45)}.
\end{alignat}
Since (P11) has the same form as (P3), it can be solved by using the method proposed in the previous part. In the following, we first explain the meaning of the symbols, and then directly give the corresponding update formulas.

Let $\bar{\rho} \textgreater 0$ denote the penalty parameter, $\boldsymbol{b}$ denote the corresponding global variables, $\{\bar{\boldsymbol{e}}_k\}_{k=1}^{K}$ denote the corresponding local variables, and $\{\bar{\boldsymbol{u}}_k\}_{k=1}^{K}$ denote the corresponding scaled dual variables. We omit the derivation steps and directly give their update formulas as follows.

\paragraph{Update $\boldsymbol{b}$}
\begin{equation}
\begin{split}
\boldsymbol{b}^{(i+1)}
&=\textrm{exp}\left(j\textrm{arg}\left(\boldsymbol{L}\hat{\boldsymbol{b}}+2\bar{\rho}\sum_{k=1}^{K}\left(\bar{\boldsymbol{u}}^{(i)}_k+\bar{\boldsymbol{e}}^{(i)}_k\right)\right)\right).
\end{split}
\end{equation}

\paragraph{Update $\{\bar{\boldsymbol{e}}_k\}_{k=1}^{K}$}
For $k=1,\dots,K$, let 
\begin{equation}
\begin{split}
\boldsymbol{\Gamma_k}&\left(\boldsymbol{b}^{(i+1)},\bar{\boldsymbol{u}}_k^{(i)}\right)=\boldsymbol{b}^{(i+1)}-\bar{\boldsymbol{u}}_k^{(i)}+\\
&\frac{\sqrt{p_k}-\left|\boldsymbol{l}_k^H \left(\boldsymbol{b}^{(i+1)}-\bar{\boldsymbol{u}}_k^{(i)}\right)\right|}{\left\|\boldsymbol{l}_k\right\|^2  \left|\boldsymbol{l}_k^H \left(\boldsymbol{b}^{(i+1)}-\bar{\boldsymbol{u}}_k^{(i)}\right)\right|}\boldsymbol{l}_k \boldsymbol{l}_k^H \left(\boldsymbol{b}^{(i+1)}-\bar{\boldsymbol{u}}_k^{(i)}\right),
\end{split}
\end{equation}
we have
\begin{equation}
\bar{\boldsymbol{e}}_k^{(i+1)}=
\begin{cases}
\boldsymbol{b}^{(i+1)}-\bar{\boldsymbol{u}}_k^{(i)};& \textrm{if} \left|\boldsymbol{l}_k^H\left(\boldsymbol{b}^{(i+1)}-\bar{\boldsymbol{u}}_k^{(i)}\right)\right|^2 \geq p_k,\\
\boldsymbol{\Gamma_k}\left(\boldsymbol{b}^{(i+1)},\bar{\boldsymbol{u}}_k^{(i)}\right) ;& \textrm{otherwise}.
\end{cases}
\end{equation}
\paragraph{Update $\{\bar{\boldsymbol{u}}_k\}_{k=1}^{K}$}
For $k=1,\dots,K$,
\begin{equation}
\bar{\boldsymbol{u}}_k^{(i+1)}=\bar{\boldsymbol{u}}_k^{(i)}+\bar{\boldsymbol{e}}_k^{(i+1)}-\boldsymbol{b}^{(i+1)}.
\end{equation}

The above formulas iterate and finally yield a suboptimal solution of (P11), denoted by $\tilde{\boldsymbol{b}}$.  Then, since $\tilde{\boldsymbol{b}}=[(t\boldsymbol{v})^T, t]^T$, the corresponding $\tilde{\boldsymbol{v}}=\left[\tilde{\boldsymbol{b}}\right]_{(1:N)} \bigg{ /}\left[\tilde{\boldsymbol{b}}\right]_{N+1}$ and $\tilde{\boldsymbol{\Psi}}=\textrm{diag}\left(\left(\tilde{\boldsymbol{v}}\right)^{\star}\right)$, where $[\boldsymbol{d}]_{(m:n)}$ denotes the vector that contains from the $m$-th element to the $n$-th element of vector $\boldsymbol{d}$.

\subsection{Overall Algorithm}

We summarize the proposed SPMC-SCA-ADMM algorithm in $Algorithm$ 3. The convergence of the proposed SPMC-SCA-ADMM algorithm can be readily shown since the objective value of (P1) is monotonically non-decreasing in the iterative process.

\addtolength{\topmargin}{0.01in}
\begin{algorithm}[h]
	\caption{SPMC-SCA-ADMM Algorithm}
	\label{alg::conjugateGradient}
	\begin{algorithmic}[1]
		\Require
		$\boldsymbol{H}_d, \boldsymbol{H}_r, \boldsymbol{G}, P$.
		\Ensure
		solution $\{\boldsymbol{x}^{op}$, $\boldsymbol{\Psi}^{op}\}$.
		\State Initialize $\hat{\boldsymbol{x}}^{(0)}$ and $\hat{\boldsymbol{v}}^{(0)}$ to feasible values, initialize $\hat{\boldsymbol{\Psi}}^{(0)}=\textrm{diag}\left(\hat{\boldsymbol{v}}^{(0)\star}\right)$, iteration number $i=0$, $\rho > 0$, $\bar{\rho} > 0$ and threshold $\epsilon > 0$.
		\Repeat
		\State Initialize $\boldsymbol{H}=\boldsymbol{H}_{r} \hat{\boldsymbol{\Psi}}^{(i)} \boldsymbol{G} + \boldsymbol{H}_{d}$ and iteration number $t_1=0$, initialize $\boldsymbol{x}^{(0)},\{\boldsymbol{e}_k^{(0)}\}_{k=1}^{K}$ and $\{\boldsymbol{u}_k^{(0)}\}_{k=1}^{K}$ to feasible values.\footnotemark[1]
		\Repeat
		\State Update $\boldsymbol{x}^{(t_1+1)}$ with formula (27).
		\State Update $\{\boldsymbol{e}_k^{(t_1+1)}\}_{k=1}^{K}$ with formula (38).
		\State Update $\{\boldsymbol{u}_k^{(t_1+1)}\}_{k=1}^{K}$ with formula (39).
		\State Update $t_1=t_1+1$.
		\Until{Convergence or the maximum number of iterations is reached.}
		\State $\hat{\boldsymbol{x}}^{(i+1)}=\boldsymbol{x}^{(t_1)}$.
		\State Initialize
		$\boldsymbol{c}_k=\textrm{diag}(\boldsymbol{h}_{r,k}^{\star})\boldsymbol{G}\hat{\boldsymbol{x}}^{(i+1)}$, $a_k=\boldsymbol{h}_{d,k}^H\hat{\boldsymbol{x}}^{(i+1)}$, $\boldsymbol{l}_k=\left[\boldsymbol{c}_k^T, a_k\right]^T$, $\boldsymbol{L}=\sum_{k=1}^K \boldsymbol{l}_k \boldsymbol{l}_k^H$ and iteration number $t_2=0$, initialize $\boldsymbol{b}^{(0)},\{\bar{\boldsymbol{e}}_k^{(0)}\}_{k=1}^{K}$ and $\{\bar{\boldsymbol{u}}_k^{(0)}\}_{k=1}^{K}$ to feasible values.\footnotemark[1]
		\Repeat
		\State Update $\boldsymbol{b}^{(t_2+1)}$ with formula (50).
		\State Update $\{\bar{\boldsymbol{e}}_k^{(t_2+1)}\}_{k=1}^{K}$ with formula (52).
		\State Update $\{\bar{\boldsymbol{u}}_k^{(t_2+1)}\}_{k=1}^{K}$ with formula (53).
		\State Update $t_2=t_2+1$.
		\Until{Convergence or the maximum number of iterations is reached.}
		\State 
		$\hat{\boldsymbol{v}}^{(i+1)}=\left[\boldsymbol{b}^{(t_2)}\right]_{(1:N)}\bigg{/}\left[\boldsymbol{b}^{(t_2)}\right]_{N+1}$ and $\hat{\boldsymbol{\Psi}}^{(i+1)}=\textrm{diag}\left(\hat{\boldsymbol{v}}^{(i+1)\star}\right)$.
		\State Update $i=i+1$.
		\Until{the fractional increase of the objective value is below the threshold $\epsilon$ or the maximum number of iterations is reached.}
	\end{algorithmic}
    \vspace{0cm}
\end{algorithm}

\footnotetext[1]{\label{initialization_method}There is a detailed description of the initialization method in \cite{huang2016consensus}.}

\section{Numerical Results}

 \begin{figure}[h]
\centerline{\includegraphics[width=0.8\columnwidth]{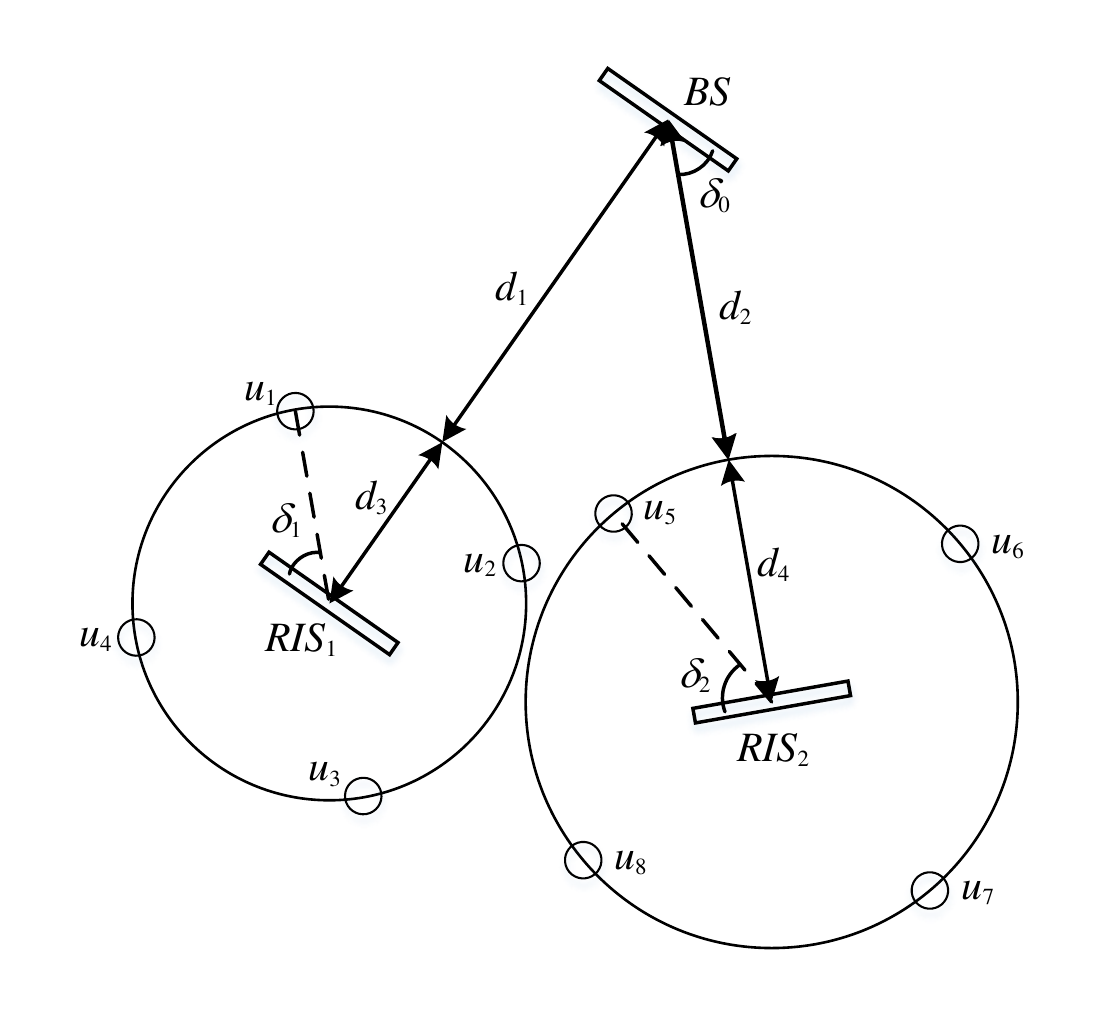}}
\caption{Simulation setup.}
\label{1}
\vspace{-0.1cm}
\end{figure}

Our simulation scenario is shown in Fig. 2, consists of one BS, two RISs and eight users. The baseband equivalent channels of both the BS-$\textrm{RIS}_1$ link and the BS-$\textrm{RIS}_2$ link are modeled as Rician fading channels whose Rician factor is $\beta_{g}$. Define user set $\mathcal{U}_1=\{u_1,u_2,u_3,u_4\}$ and user set $\mathcal{U}_2=\{u_5,u_6,u_7,u_8\}$. Then, the baseband equivalent channels of the the links from $\textrm{RIS}_i$ to users in $\mathcal{U}_i$ ($i$=1, 2) are also modeled as Rician fading channels, whose Rician factor is $\beta_{hr}$. In the simulation, we set $\beta_{g}=2$ and $\beta_{hr}=2$. As for the baseband equivalent channels of the BS-users link and the links from $\textrm{RIS}_i$ to users in $\mathcal{U}_j$ with $i,j\in \{1,2\}$, $i\not=j$, they are modeled as Rayleigh fading channels since we consider that there are obstacles on these links. We set the passloss factor $n=3$ for all the baseband equivalent channels. Assume the geometric size of the two RISs is much larger than the wavelength, so that RISs can be modeled as specular reflectors \cite{basar2019wireless}. Further, we consider uniform linear arrays (ULAs) at the BS and the two RISs. Specifically, $\textrm{RIS}_1$ is positioned in parallel to the BS antenna array, and we set $\delta_0=\pi /4$, $\delta_1=\pi /4$, $\delta_2=\pi /3$. Other system parameters are given as: carrier frequency $f_c=755\textrm{MHz}$, $P = 10 \textrm{W}$, $K=8$, $d_1=8\textrm{m},d_2=7\textrm{m}$, $d_3=4\textrm{m}$ and $d_4=5\textrm{m}$.

\begin{figure}[htb]
	\centerline{\includegraphics[width=0.9\columnwidth]{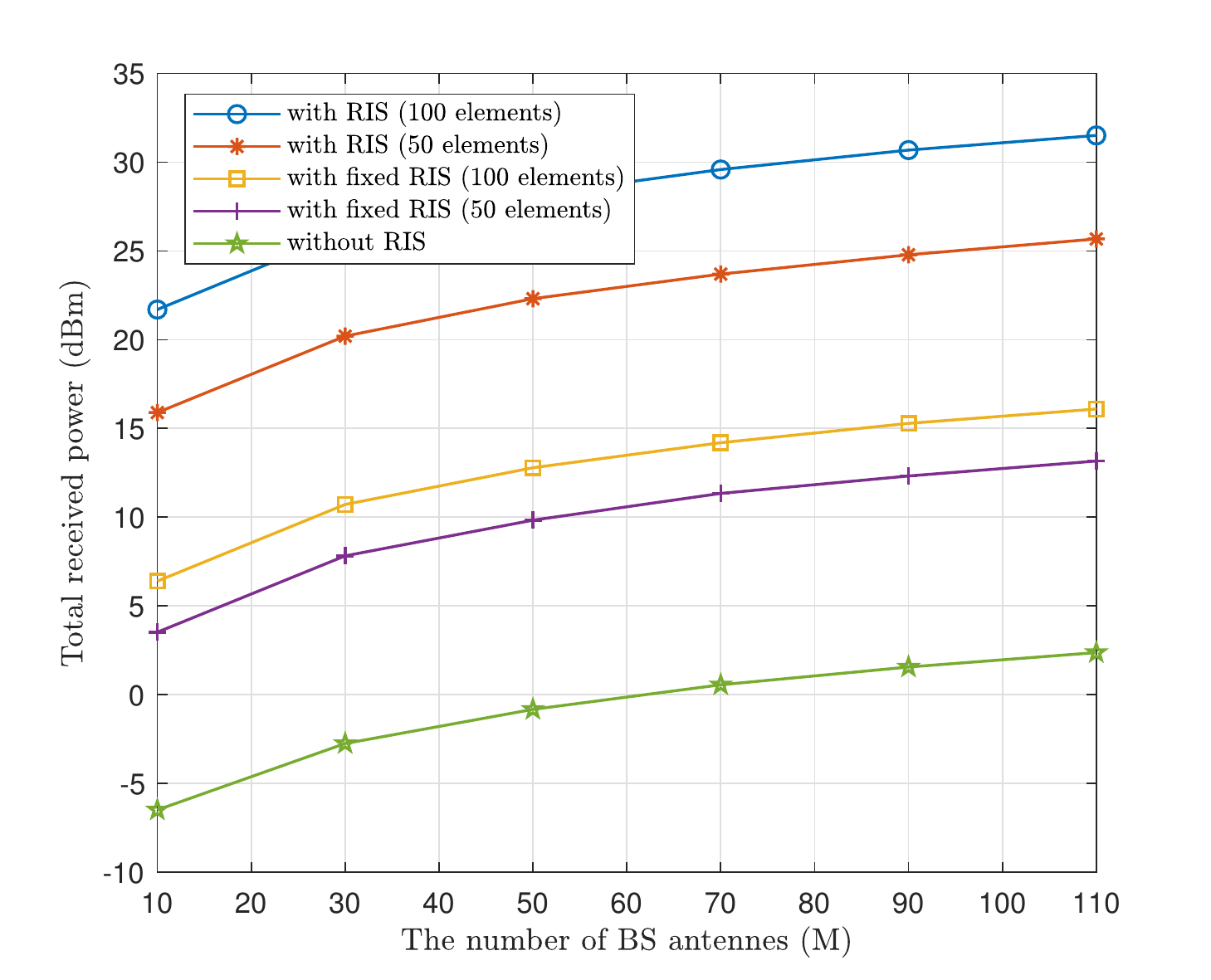}}
	\caption{Total received power versus the number of BS antennes, M.}
	\label{2}
	\vspace{-0.4cm}
\end{figure}

Fig. 3 shows the total received power versus the number of BS antennes $M$ for different states of RIS. To make sure that (P1) is solvable, we set all the elements in $\left\{p_k\right\}_{k=1}^K$ to be equal to $0$. Naturally, in the figure, as $M$ increases, the total received power increases. Further, the figure shows the total received power under different RIS states. In the absence of RIS, the total received power is the smallest, while in the presence of RIS, the larger $N$ is, the greater the total received power will be. It is worth noting that as long as the RISs are placed in the environment, the total received power will still increase even if the RISs are not optimized. This is true since placing RIS in the environment introduces new energy transfer paths.



\begin{figure}[htb]
	\centerline{
	\subfigure[Minimun received power]{
		\begin{minipage}[t]{0.45\columnwidth}			
			\centerline{\includegraphics[width=\columnwidth]{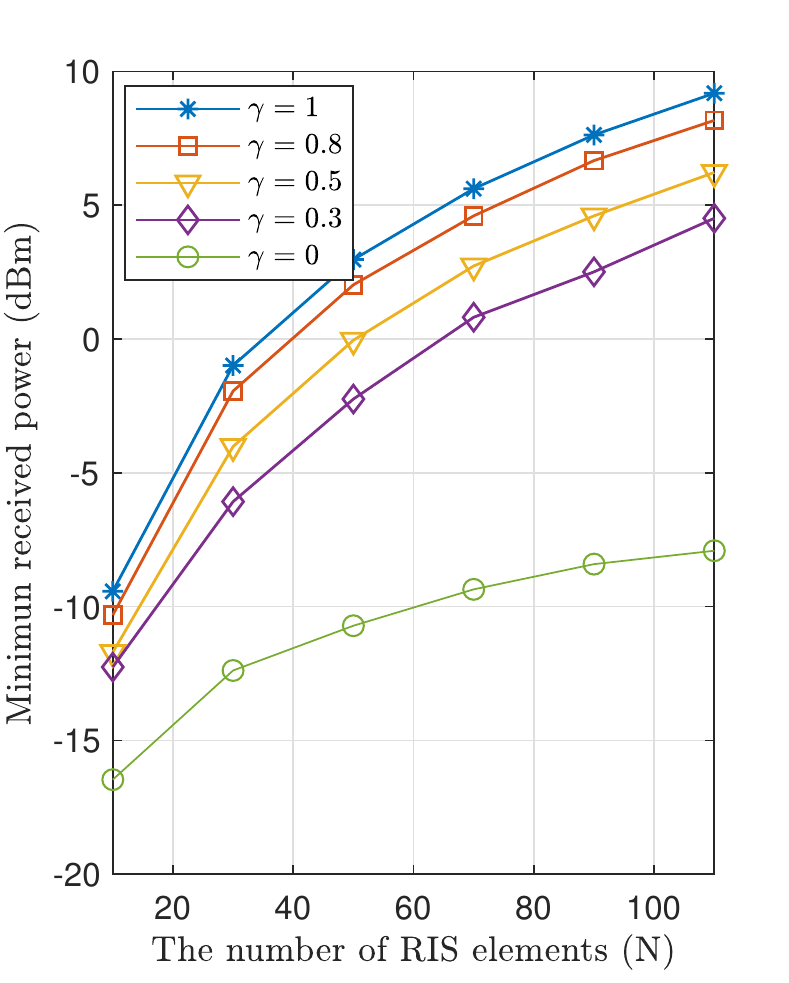}}
		\end{minipage}%
	}%
	\subfigure[Total received power]{
		\begin{minipage}[t]{0.45\columnwidth}
			\centerline{\includegraphics[width=\columnwidth]{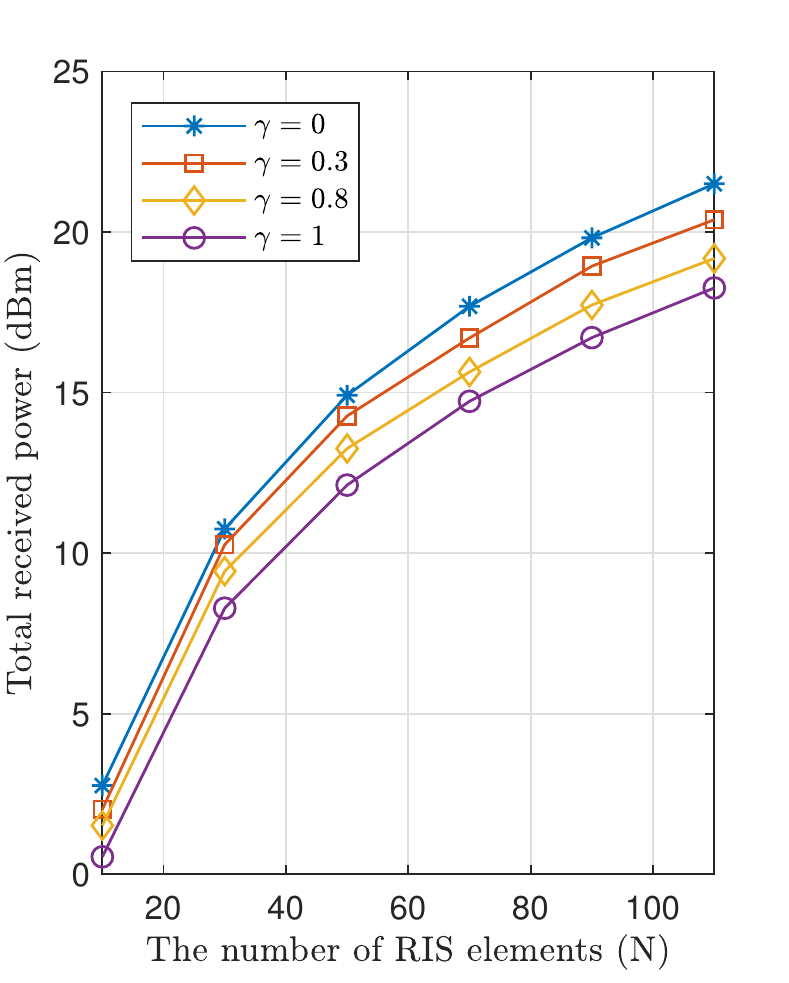}}
		\end{minipage}%
	}}
	\caption{Simulation results with different $\gamma$.}
	\vspace{-0cm}
\end{figure}

Fig. 4 (a) and (b) show our simulation results with different $\gamma$. Note that we set all of the elements in $\left\{p_k\right\}_{k=1}^K$ to be equal to $\gamma Q_{MM}$ in this simulation, where $\gamma \in [0,1]$ is a scaling factor and $Q_{MM}=\max_{\boldsymbol{\Psi}, \boldsymbol{x}} \min_{1,\dots,K} Q_k$. It is seen that with this setup, (P1) is solvable. In the simulation, we obtain the value of $Q_{MM}$ under different channel states by trial and error. As seen from this two figures, with the increase of $N$, both the total received power and the minimum received power of all the $K$ users improve. And with the increase of $\gamma$, the total received power decreases while the minimum received power increases. This makes sense since a larger $\gamma$ means tighter constraints. For further explanation, a larger $\gamma$ means a stronger QoS requirement, corresponding to a higher minimum received power in Fig. 4 (a), while a stronger QoS requirement leads to a reduction of energy efficiency, corresponding to a smaller total received power in Fig. 4 (b).

\section{Conclusion}

In this paper, we propose a novel scheme to improve the energy efficiency of the RIS-aided wireless power transfer system. To balance energy efficiency and user fairness, we formulate a problem to maximize the total received power of all the users by jointly optimizing the beamformer at transmitter and the phase shifts at the RISs, subject to the minimum received power constraints. We design a low-complexity algorithm for this problem by applying alternating optimization techniques, SCA method and ADMM method. Numerical results demonstrate the effectiveness of the proposed algorithm and show the trade-off between energy efficiency and user fairness.




\ifCLASSOPTIONcaptionsoff
  \newpage
\fi



%

\bibliographystyle{ieeetr}

\bibliography{ref}




%








\end{document}